\documentclass{revtex4}

\usepackage{graphicx}
\usepackage{dcolumn}
\usepackage{bm}
\usepackage{epstopdf}
\usepackage{multirow} 
\usepackage{hyperref}
\usepackage{color}
\begin{document}
\preprint{APS/123-QED}

\title{Theoretical description of X-ray absorption spectroscopy of the graphene-metal interfaces}

\author{Elena Voloshina,$^{1,}$\footnote{Author to whom correspondence should be addressed. Electronic mail: elena.voloshina@hu-berlin.de} Roman Ovcharenko,$^{2}$ Alexander Shulakov,$^{2}$ and Yuriy Dedkov$^{3}$}
\affiliation{$^{1}$Humboldt-Universit\"at zu Berlin, Institut f\"ur Chemie, 12489 Berlin, Germany\\
$^{2}$V.~A.~Fock Institute of Physics, Saint-Petersburg State University, 198405 St. Petersburg, Russia\\
$^{3}$SPECS Surface Nano Analysis GmbH, Voltastra\ss e 5, 13355 Berlin, Germany
}

\date{\today}

\begin{abstract}

The present manuscript considers the application of the method of the near-edge X-ray absorption spectroscopy (NEXAFS) for the investigation of the graphene-based systems (from free-standing graphene to the metal-intercalation-like systems). The \mbox{NEXAFS} spectra for the selected systems are calculated in the framework of the approach, which includes the effects of the dynamic core-hole screening. The presented spectral changes from system to system are analysed with the help of the corresponding band-structure calculations. The obtained results are compared with available experimental data demonstrating the excellent agreement between theory and experiment. The direct correlation between the strength of the graphene interaction with the metallic substrate and the spectral distributions (shape and intensities of $\pi^*$ and $\sigma^*$ features in the C $K$ NEXAFS spectra) is found that can be taken as a fingerprint for the description of interaction at the graphene/metal interface.
 
\end{abstract}

NEXAFS/XES/XMCD project: {\color{blue}\underline{\url{https://sites.google.com/site/elsaspectroscopy/}}}

\maketitle

\section{Introduction}

\begin{figure}[t]
\centering
\includegraphics[width=0.42\textwidth]{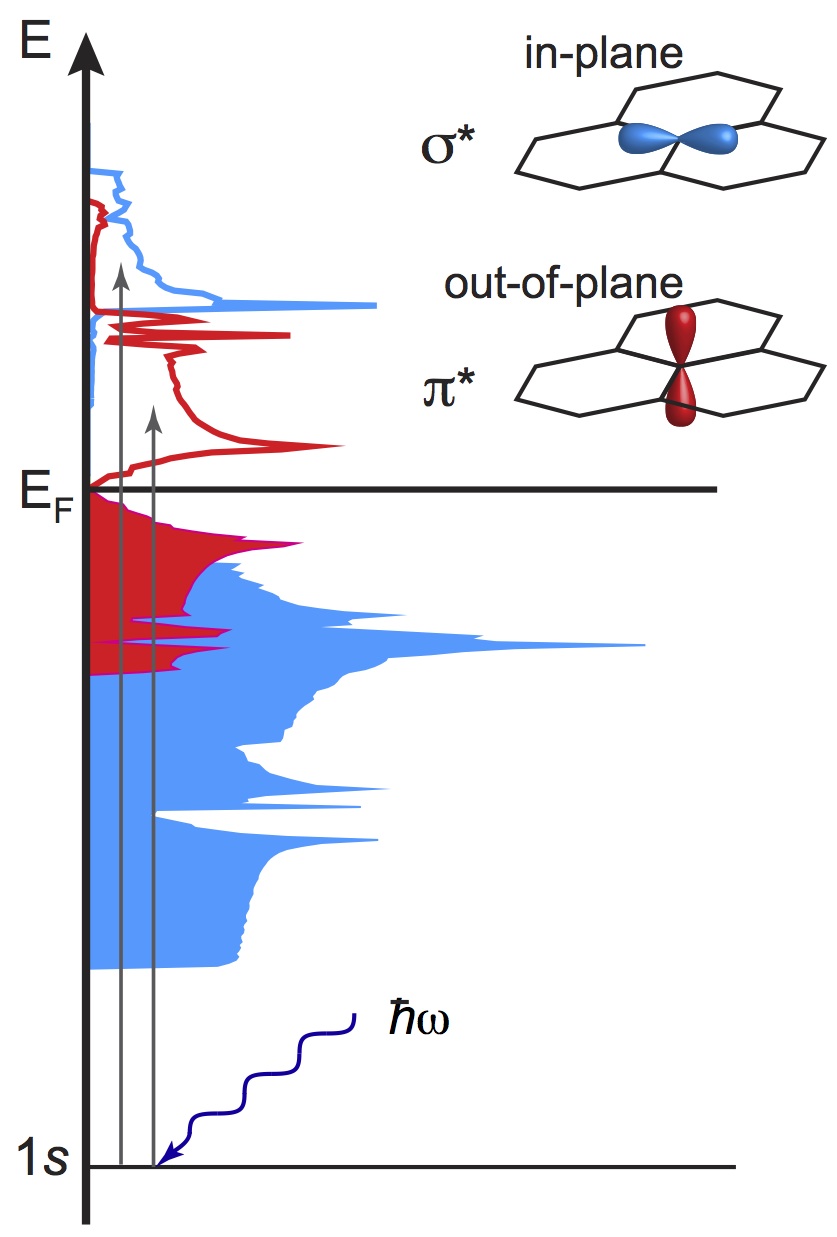}
\caption{\label{xas-scheme} The scheme of the X-ray light absorption process. The light from the synchrotron light source is scanned around the energy corresponding to the binding energy of the particular core level and in this case the absorption coefficient is proportional to the density of the valence band states above $E_F$. The orbital selectivity is reached via relative orientation of the linearly polarised light and the sample.}
\label{fig:scheme}
\end{figure}

The method of x-ray absorption spectroscopy (XAS) is one of the widely used spectroscopic tools for electronic structure investigations in condensed matter physics~\cite{Stohr:1992NEXAFSbook}. It is based on the absorption of the X-ray photons when core shell electron is transferred into the unoccupied valence band states (conduction band) above the Fermi level ($E_F$) (Fig.~\ref{fig:scheme}). Therefore one or more jumps (absorption edges) are usually observed in the absorption spectrum. Moreover the energy position of any edge is element specific since it coincides with the energy of the corresponding atomic core level. Besides, X-ray transitions are controlled by the dipolar selection rules and for well-defined atomic symmetry of core hole the final state angular momenta are selected in XAS. Strong spatial localisation of the initial core shell state leads also to the site specific behaviour of XAS. 

Generally, appearance of the fine structure of XAS of solids can be described in the framework of photoelectron scattering model. Near the edge photoelectron energy is low and photoelectron de Broglie wave undergoes multiple scattering on neighbouring atoms. This leads to the origin of the reach and intense fine structure, or near edge X-ray absorption fine structure, NEXAFS. On the other hand, in the large photoelectron energies region (far from the XAS edge) scattering of the photoelectron waves gives rise to rather simple and weak variation of absorption coefficient, or extended X-ray absorption fine structure, EXAFS, well described by single scattering theory. It is considered that the range of NEXAFS covers up to 50 eV above the absorption edge and the rest high photon energy part of XAS extended up to 1000 eV above the threshold belongs to EXAFS. The latter method is very effective tool for the short range atomic ordering determination in multiatomic systems including noncrystalline solids. In NEXAFS spectroscopy, the topic of the present paper, absorption spectra can be interpreted on the basis of the corresponding molecular orbital or band-structure calculations above $E_F$. This way is equivalent to the interpretation of XAS data in the frame of photoelectron waves multiple scattering model, transparent and demonstrative to date.

To measure spectral distribution of the X-ray absorption coefficient in massive (bulk) solids yield spectroscopy is usually used -- the dependence on photon energy of the total photocurrent (so called total electron yield, TEY). Since external photo-effect bunch, which accompanied X-ray absorption event, consists of primary fast photo- and Auger-electrons together with the cloud of slow secondary electrons, the mean drift way of the whole electron bunch is govern by the relatively large mean free path of slow secondary electrons~\cite{Frazer:2003a}. In order to increase surface sensitivity of measurements (for the better representation of the surface-related electronic structure) the negative repulsive potential can be applied to the grid-electrode placed in front of the detector to suppress slow kinetic energy component in the electron bunch distribution. This experimental mode gives so-called partial electron yield (PEY). In the case of PEY NEXAFS with the linearly polarized light, the method can be used for the determination of the orbitals' spatial orientation. Generally, one can observe the so-called \textit{search-light-like effect}~\cite{Stohr:1999b}, which can be used for probing the quadrupole moment of the local charge around the absorbing atom. In such XAS experiment, the absorption intensity associated with a specific molecular orbital final state has a maximum if the electric field vector is aligned parallel to the direction of maximum charge or hole density, i.\,e. along a molecular orbital, and the intensity vanishes if the electric field vector is perpendicular to the orbital axis. A detailed description of the angular dependence of NEXAFS intensities can be found elsewhere~\cite{Stohr:1999b,Stohr:1999a}. In case of magnetic samples and circularly polarized light the absorption coefficient depends on the relative orientation of photon spin and magnetization direction of the sample. The quantitative analysis of the absorption spectra obtained on the magnetic sample with the circularly polarized X-rays can be performed with the help of the so-called magneto-optical sum rules for evaluation of spin- and orbital-magnetic moments~\cite{Thole:1992,Carra:1993}.

Compared to the relative simplicity of the NEXAFS experiment, the theoretical description of X-ray absorption process in solid is not that easy. The main problem here is that the excited system with the core hole has to be adequately described, which is always a difficult task for the standard density functional theory (DFT) calculations. Particularly it is related to graphite and graphene-based systems.

The NEXAFS spectrum of graphite or graphene can be separated on two energy regions (see Fig.~\ref{fig:scheme}): (i) $1s\rightarrow\pi^*$ contribution originating from transition of $1s$ core electron on C\,$p_z$ orbitals and (ii) $1s\rightarrow\sigma^*$ one originating from transition of $1s$ core electron on C\,$sp^2$ hybrid orbitals~\cite{Bruhwiler:1995,Pacile:2008,Hua:2010}. As was recently shown~\cite{Wessely:2005}, although the $\pi^*$ resonance was well reproduced in the framework of the core-hole final state approximation, the shape as well as the intensity of the $\sigma^*$ resonance were not reproduced. Only the application of the Mahan-Nozier\'es-De Dominicis (MND) theory of dynamical core-hole screening gives an adequate description of the NEXAFS spectrum of graphite~\cite{Wessely:2005,Wessely:2006}. It has to be noted here that in the double-peak structure of the $\sigma^*$ resonance, the first peak ($\sigma^*_1$), which reflects the excitonic final state can be reproduced in the final state and MND approaches, and second $\sigma^*_2$ peak is described as due to the quickly completely screened initial-state density of states. 

The present manuscript is devoted to the spectroscopic studies of the graphene-metal systems. The properties of these objects are the subject of many experimental and theoretical studies which were recently reviewed in several works~\cite{Wintterlin:2009,Batzill:2012,Dedkov:2012book,Voloshina:2012c}. As discussed the graphene/metal interface is an important object of the graphene-related fields of research. Beyond the practical utilisation of these objects (ideal ways to prepare the huge layers of graphene, graphene-based spin filters and gas sensors, etc.)~\cite{Geim:2009,Kim:2009a,Bae:2010,Karpan:2007,Dedkov:2010a}, there is also a fundamental interest to the nature of the interaction between graphene and metal~\cite{Wintterlin:2009,Voloshina:2012c}. As the NEXAFS spectroscopy can be used for the investigation of the empty electronic states of material in the vicinity of $E_F$, it can be utilised as a complimentary to photoemission spectroscopy method, which can shed more light on the nature of interaction between graphene and a metal.  

Here we revisited the theoretical description of the absorption spectroscopy for graphene-based systems. We used the scheme described in Refs.~\cite{Wessely:2005,Wessely:2006} and our computer code for the modelling of NEXAFS spectra of the graphene-based systems: graphene single- and bilayer on Ni(111) and graphene/Al/Ni(111) intercalation-like system. Our results, starting from the test example (graphene) and for more complex systems, are discussed on the basis of effects in the electronic structure of the studied objects and compared with available experimental results. Very good agreement between theory and experiment is found in our analysis.

\section{Theoretical and Computational details}
\label{comput}

\subsection{Simulation of the X-ray absorption spectra}

The X-ray absorption spectroscopy, being one of the oldest and most efficient spectroscopic tools to investigate the electronic structure of materials, involves a very complicated dynamical many-body process, so that an adequate theoretical treatment of the experimental data appears to be a very complicated problem. When the X-ray photon is absorbed, the electron from the inner shell excites into the unoccupied states of the conduction band where electronic states are described by Wannier functions localized on the absorbed atom. The contribution of transitions into the states centred at neighbouring atoms is considered to be small enough to be neglected.

There are three main approaches for the description the X-ray absorption process. Generally accepted are first two one-electron ways of NEXAFS interpretation in terms of the initial state or the final state of system during X-ray absorption. Less common is the third way in which many-body processes tracked the X-ray transitions are taking into account. These ways are different one of other sonly in approximations of core-hole screening:
\begin{itemize}
\item The first approach assumes that the time of the interaction of the X-ray photon with the electronic subsystem is much shorter than any typical process in the system (initial-state approximation).
\item The second approach assumes that the photon-electron interaction is adiabatically slow that the system has a time for transfer from the ground state into a new one (final-state approximation).
\item The MND theory assumes the creation of the core-hole as fast enough process, but takes into account the real dynamics of the response of the electronic subsystem to a core hole creation (MND approximation).  
\end{itemize}  

\subsubsection{One-electron approximations. }

In the initial-state approach, the first principle calculations of the electronic structure of material are performed with VASP code (for details, see Sec.~\ref{dft}). Then, we obtain the projections of the Bloch wave function onto the spherical harmonics centred on site $\alpha$ (atomic number $Z$) and construct the matrix of initial-state Green's functions as following:
\begin{equation}
g_{mn}(\omega)=\sum_{\mathbf{k},l}\frac{<\alpha,m|\mathbf{k},l><\mathbf{k},l|\alpha,n>}{\omega-E_{\mathbf{k},l}-\mu+i\delta},
\label{g-mn}
\end{equation}
where $m$, $n$ are angular momentum for ($s$, $p_1$, $p_0$, and $p_{-1}$) orbitals, $\mathbf{k}$ - wave vector, $l$ - band index, $E$ - energy, $\mu$ - chemical potential, and $\delta$ - infinitely small value. Then the initial-state NEXAFS spectrum is given by equation:
\begin{equation}
I^i(\omega)\propto\sum_{m,n}t_m\mathrm{Im}\,g_{mn}t^*_n,
\label{I}
\end{equation}
where $t_n=\mathbf{E}<1s|\mathbf{p}|n>$, $\mathbf{p}$ is the dipole moment operator, and $\mathbf{E}$ - X-ray electric field. 

In our work we used these approximations as follows. We model the studied system with a large supercell [i.e. $(6\times6)$ in the case of free-standing graphene]. Then we replace one atom by the following atom in the Periodic Table - the so-called $Z+1$ approximation. The density of states (DOSs) are obtained by means of VASP calculations performed for the large supercell and the Bloch wave functions are projected onto spherical harmonics centred on the $Z+1$-atom site to get an expression for the final state, and on the $Z$-atom site to get an expression for the initial state.

The final-state NEXAFS spectrum is calculated according to the equation similar to eqn. (\ref{I}), where the Bloch wave functions are projected onto atom $\beta$ with an atomic number $(Z+1)$, and the final-state Green's function $G_{mn}(\omega)$ is written in the same manner as eqn. (\ref{g-mn}).

\subsubsection{MND approximation. }

\begin{figure}
\centering
\includegraphics[width=0.92\textwidth]{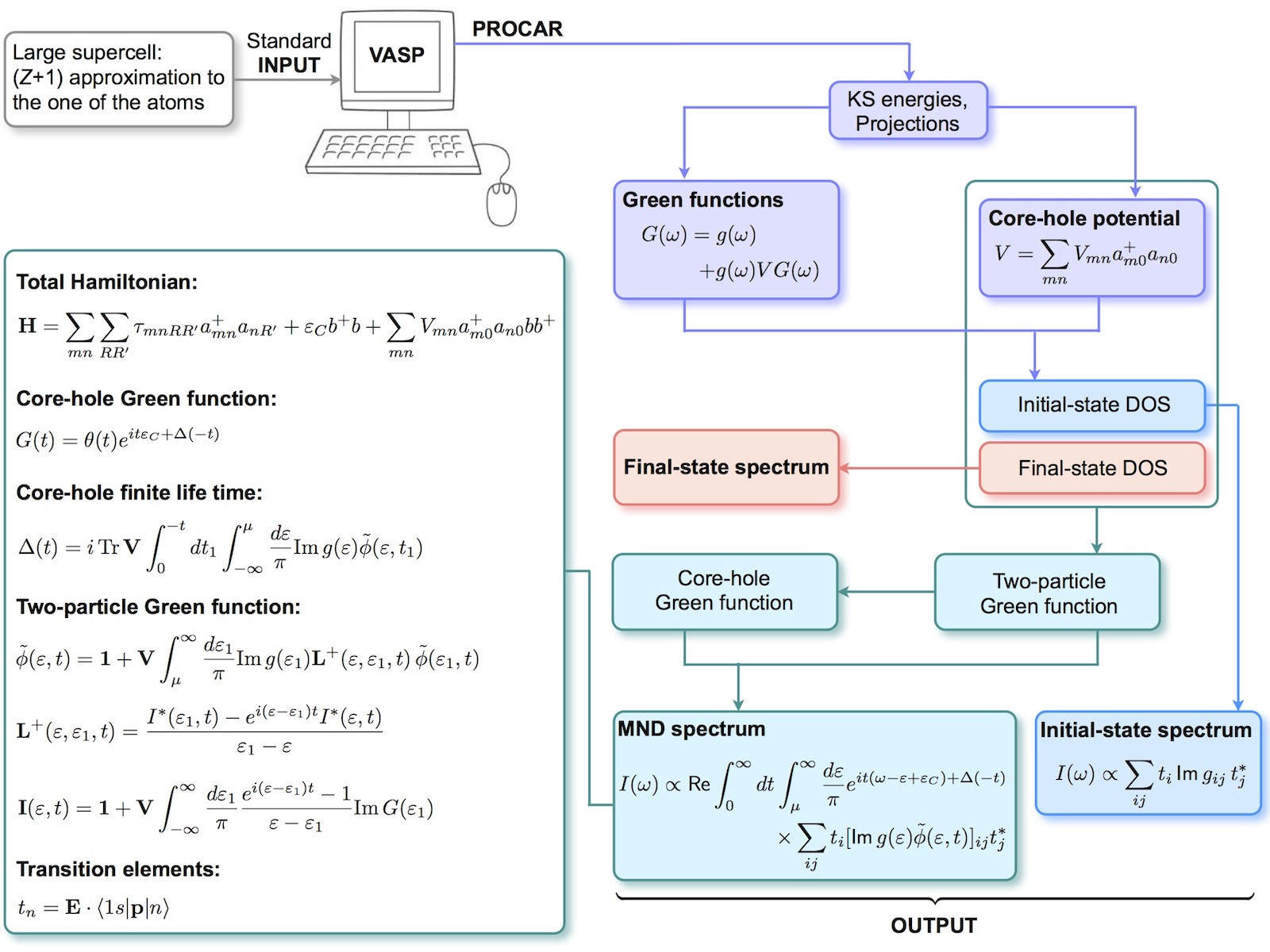}
\caption{Computational algorithm used for calculation of NEXAFS spectra.}
\label{fig:xas-scheme}
\end{figure}

The MND theory considers the creation of core-hole as an instant process taking into account real dynamics of the response of the electronic subsystems. As a result many-electrons effects due to correlations of electrons with core-hole are included. Therefore we construct initial- and final-state Green's functions as mentioned above and the so-called core-potential:
\begin{eqnarray}
V_{mn}&=&\sum_{\mathbf{k},l}(<\beta,m|\mathbf{k},l><\mathbf{k},l|\beta,n>-%\\
 %& &
 <\alpha,m|\mathbf{k},l><\mathbf{k},l|\alpha,n>).\nonumber
\end{eqnarray}
The MND spectrum is calculated according to the formula shown in the lower part of Fig.~\ref{fig:xas-scheme}. The formulas for the calculation of the two-particles Green's function $[\tilde\phi(E,t)]$ and core-hole finite lifetime $[\Delta(t)]$ are presented in the left-hand side of the scheme in Fig.~\ref{fig:xas-scheme}. The full description of the calculation algorithm can be found in Refs.~\cite{Wessely:2005,Wessely:2006} and in the Supplementary material~\cite{suppl}.

\subsection{DFT calculations}
\label{dft}

The DFT calculations were carried out using the projector augmented wave method~\cite{Blochl:1994}, a plane wave basis set and the generalized gradient approximation as parameterized by Perdew \textit{et al.}~\cite{Perdew:1996}, as implemented in the VASP program~\cite{Kresse:1994}. The plane wave kinetic energy cutoff was set to $500$\,eV. The long-range van der Waals interactions were accounted for by means of a semiempirical DFT-D2 approach proposed by Grimme~\cite{Grimme:2010}. The corresponding structures of the graphene-metal based systems are shown in Fig.~\ref{fig:structure} and they are discussed in details further in the text. The supercell used to model the graphene-metal interface is constructed from a slab of $13$ layers of nickel atoms with a graphene layer (or bilayer) adsorbed at both sides and a vacuum region of approximately $18$\,\AA. In the case of graphene/Al/Ni(111) intercalation-like system, the supercell has $(2\times2)$ lateral periodicity and contains by one Al layer (3 atoms each) introduced between graphene and Ni(111) on both sides of the slab. In optimizing the geometry, the positions ($z$-coordinates) of the carbon atoms as well as those of the top two layers of metal atoms are allowed to relax. In the total energy calculations and during the structural relaxations the $k$-meshes for sampling the supercell Brillouin zone are chosen to be as dense as $24\times24$ and $12\times12$, respectively. 

To obtain an input for the NEXAFS spectra simulations, we employed large supercells of $(6\times 6)$ and $(4\times4)$ periodicity when studying free-standing graphene and graphene/metal systems, respectively. In the latter case the number of nickel layers was reduced to $5$. In order to have the same position for the Fermi level, both, initial and final states DOSs, are taken from the same calculation sets.

\section{Results and Discussion}

In this part we present the results of the application of the computational procedure discussed in the previous section for the studies of a graphene layer on metallic support. We consider several objects in which the strength of interaction of graphene with metal is ranged from ``strong'' [graphene/Ni(111)] to ``weak'' [graphene/Al/Ni(111) and graphene-bilayer/Ni(111)]. 

\subsection{Free-standing graphene vs graphene/Ni(111)}

\begin{figure}
\centering
  \includegraphics[width=0.48\textwidth]{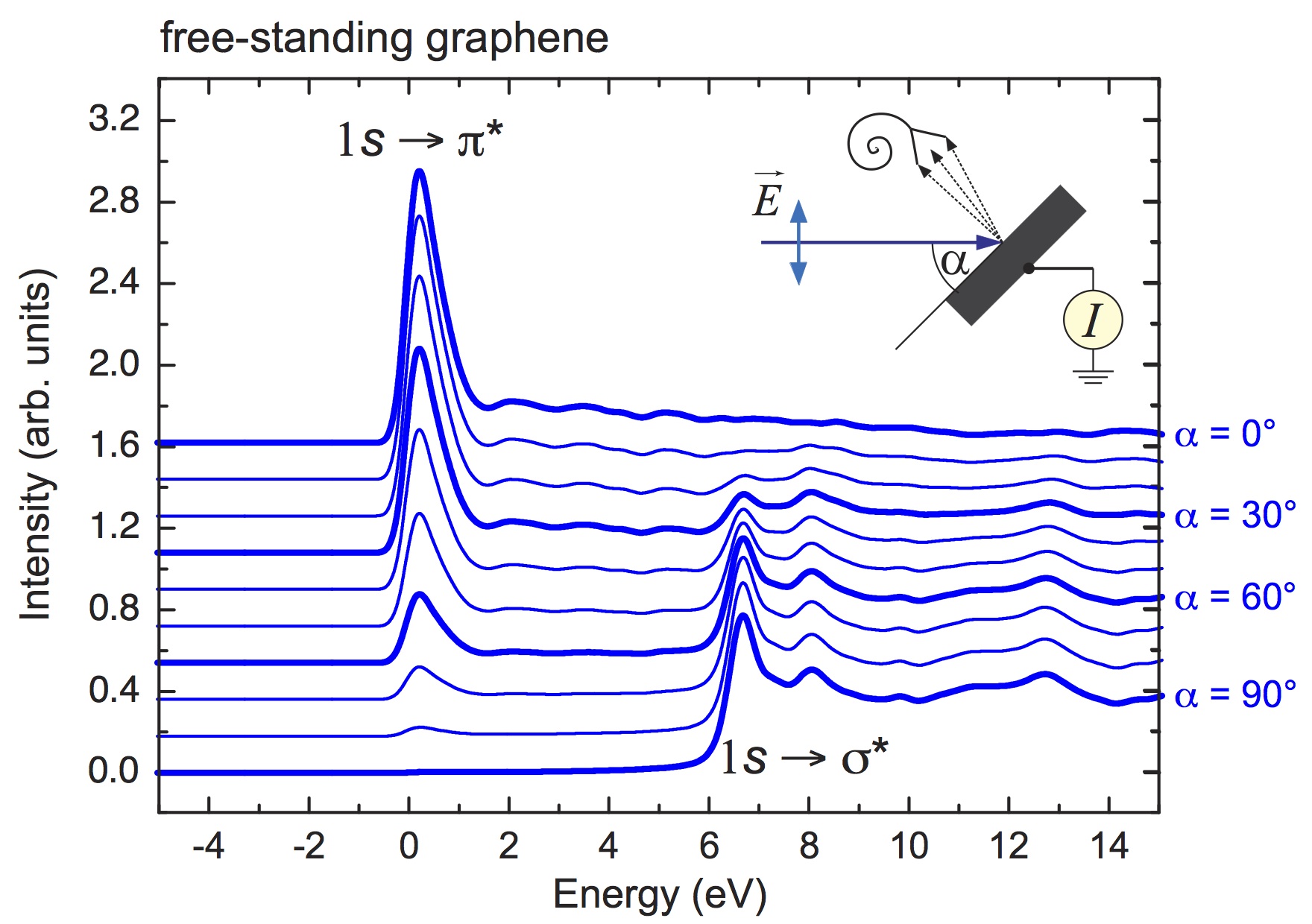}
  \caption{Calculated, in a framework of MND approach, C $K$ NEXAFS spectra for a free-standing graphene as a function of the angle, $\alpha$, between electric field vector of X-ray and the normal to the surface. The experimental geometry is shown as an inset.}
  \label{fig:FSG}
\end{figure}

\begin{figure}
\centering
  \includegraphics[width=0.48\textwidth]{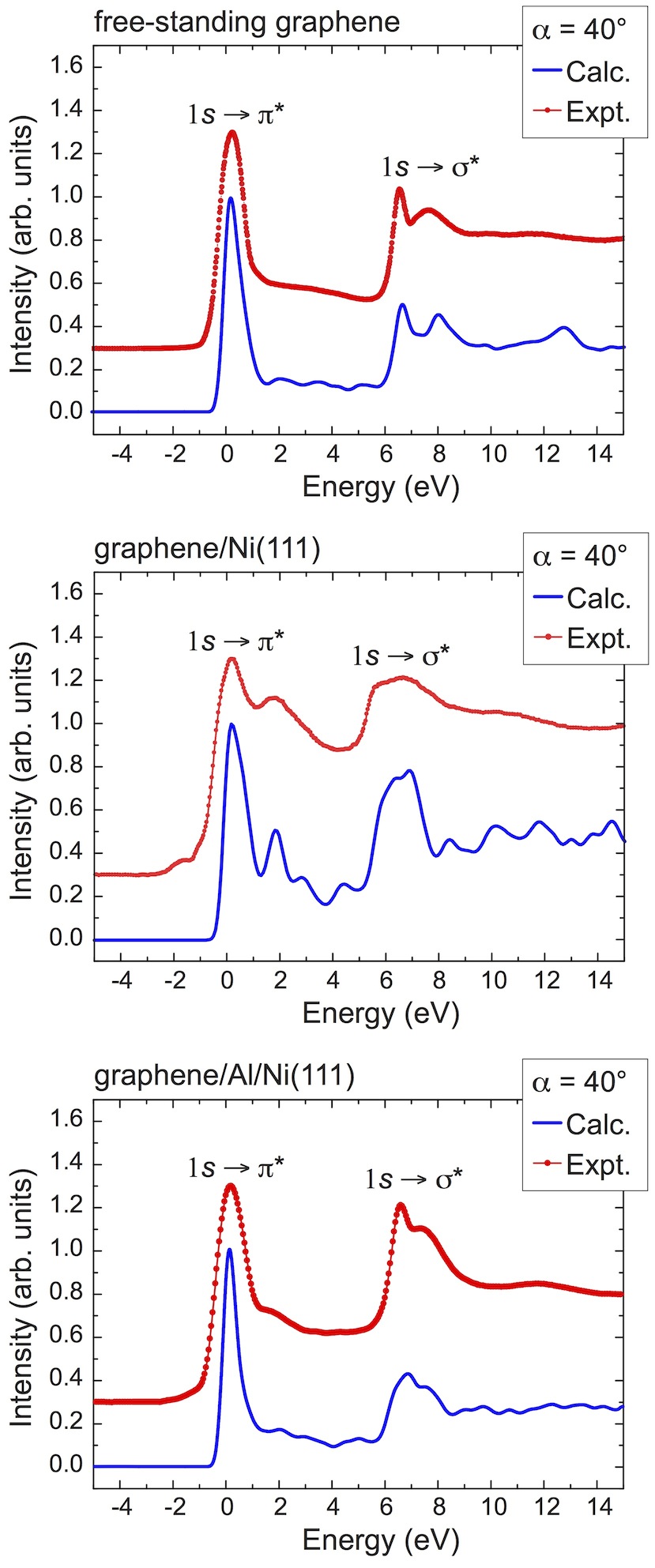}
  \caption{Comparison of experimental and calculated NEXAFS spectra obtained with $\alpha = 40^\circ$ at the C $K$ absorption edge for graphite, graphene/Ni(111), and graphene/Al/Ni(111). Experimental data are taken from Refs.~\cite{Weser:2010} and \cite{Voloshina:2011NJP}.}
  \label{fig:theoryvsexp}
\end{figure}

In order to prove our implementation of the algorithm discussed earlier we performed the calculation of NEXAFS spectra for the free standing graphene (Figs.~\ref{fig:FSG} and \ref{fig:theoryvsexp}). The main peaks of the graphene spectrum at $1.9$\,eV and $8.3$\,eV above $E_F$ correspond to transitions of the $1s$ core electron onto the $\pi^*$ and $\sigma^*$ states, respectively. Fig.~\ref{fig:FSG} shows the theoretical spectra calculated according to the procedure described in Sec.~\ref{comput} for ten different incidence angles $\alpha$. One can note a smooth decreasing (increasing) weights of $1s \rightarrow \pi^*$ ($1s \rightarrow \sigma^*$) transitions when going from $\alpha=0^\circ$ to $\alpha=90^\circ$, that is in accordance with geometry of the absorption process. This example is a nice demonstration of the \textit{search-light-like} effect discussed earlier: for the grazing incident angles the maximum of absorption is observed for $\pi^*$ states and for angles close to the normal to the surface the maximum of absorption is detected for $1s \rightarrow \sigma^*$ transitions. Furthermore, it is clearly seen that both the shape and positions as well as relative intensities of the peaks are well reproduced by theory [for comparison between calculated and experimental spectra~\cite{Weser:2010,Dedkov:2010a}, see Fig.~\ref{fig:theoryvsexp} (upper panel)]. As was discussed earlier in Refs.~\cite{Wessely:2005,Wessely:2006} the shape as well as the position of $1s\rightarrow \pi$ and $1s\rightarrow \sigma^*_1$ transitions are well reproduced by the MND theory described in Fig.~\ref{fig:xas-scheme}. The second peak $\sigma^*_2$ can be reproduced only if fully screened initial states density of states is taken into account and the corresponding initial states NEXAFS spectrum is summed with MND spectrum in order to reproduce the experimental observations.

Before investigating modifications in the absorption spectra due to the presence of the Ni(111) substrate underneath graphene, let us consider briefly the atomic structure of the interface. The surface lattice constant of Ni(111) matches the lattice constant of graphene almost perfectly and therefore $(1\times 1)$ structure is formed. The structure of graphene/Ni(111) widely accepted in the literature has the carbon atoms arranged in the so-called \textit{top-fcc} configuration on Ni(111): one carbon atom from the graphene unit cell is placed above the Ni interface atom and the second one occupies the hollow \textit{fcc} site of the Ni(111) surface (Fig.~\ref{fig:structure}). Our calculations support this model (see Supplementary Tab. S1~\cite{suppl}). 

The equilibrium distance between graphene layer and Ni(111) is relatively small, indicating ``strongly'' bonded system: significant hybridization of the valence band states of graphene and Ni leads to the complete destruction of the graphene Dirac cone~\cite{Bertoni:2004,Karpan:2008,Weser:2011,Voloshina:2012c}. [This effect of hybridization is clearly seen when considering electronic band structure of the system (see Supplementary Fig. S1~\cite{suppl}). The modifications in the electronic structure are considered in details in Ref.~\cite{Bertoni:2004}].

Consequently, the NEXAFS spectra in this case are modified with respect to those of free-standing graphene [Fig.\ref{fig:theoryvsexp} (middle panel)]. Firstly, in the region of the $1s \rightarrow \pi^*$ transition the graphene/Ni(111) absorption spectra have a double-peak structure compared to those of graphite that is explained by the transitions of C $1s$ core electron into two unoccupied states (interface states), which are the result of hybridization of the graphene $\pi$ and Ni $3d$ valence band states. The first peak at about $0.2$\,eV corresponds to the interface state which is allocated in the range between $0$ and $1$\,eV above $E_F$ around $K$-point of the Brillouin zone, having slightly different energy depending on the type of C-atom ($\mathrm{C}_\mathrm{Ni}^{top}$ or $\mathrm{C}_\mathrm{Ni}^{fcc}$) and on the spin-orientation. The second peak, near $2$\,eV, comes mainly from the unoccupied interface state at the $M$ point (see Supplementary Figs. S1 and S2~\cite{suppl}). From this analysis one can clearly see that both peaks belong to the hybrid states where both carbon atoms are involved. Therefore the previous analysis of NEXAFS spectra for the graphene/Ni(111) system performed in Refs.~\cite{Weser:2010,Dedkov:2010a,Rusz:2010} on the basis of calculations of the $K$-edge electron loss spectra~\cite{Bertoni:2004} is not fully correct (the double peak structure was assigned to different carbon atoms in the unit cell). Secondly, the visible reduction in the energy separation between the $\pi^*$ and $\sigma^*$ features are observed for all graphene/Ni(111) spectra compared to that in the spectra of free-standing graphene, in accordance with the shift of the $\sigma^*$ state for the carbon atoms at the interface by ca.\,$1$\,eV to lower energies that can be explained by the lateral bond softening within the adsorbed graphene monolayer and can be connected with the downward shift of $\sigma^*$ states for graphene/Ni(111) compared to free-standing graphene. %The similar results were recently obtained in Ref.~\cite{Rusz:2010}.

\subsection{Decoupling graphene from the substrate: The gra\-phe\-ne/Al/Ni(111) system}

\begin{figure}
\centering
  \includegraphics[width=0.92\textwidth]{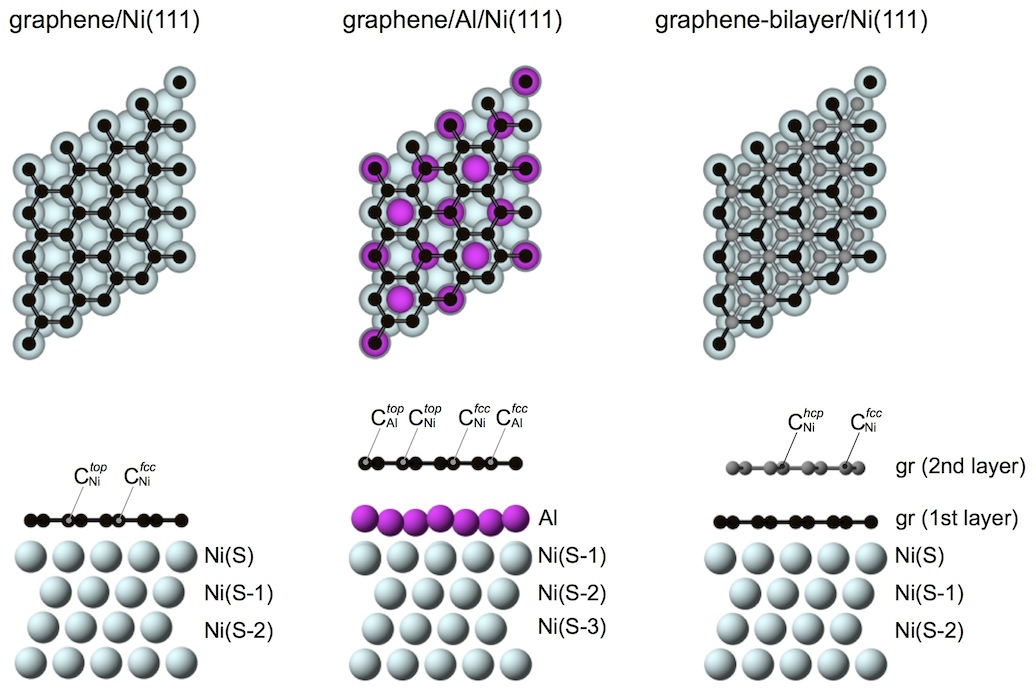}
  \caption{Top- and side-views of the crystallographic structures of the studied systems: graphene/Ni(111), graphene/Al/Ni(111), and graphene-bilayer/Ni(111) (from left to right).}
  \label{fig:structure}
\end{figure}

In the graphene/Al/Ni(111) system, the \textit{fcc} Al(111) monolayer is placed in the space between graphene and Ni(111) (Fig.~\ref{fig:structure}). There is only one possible arrangement of aluminium atoms in the high-symmetry positions as shown in the figure: the Al(111) lattice plane is rotated by $30^\circ$ with respect to the graphene/N(111) lattice such that the Al atoms occupy all three kind of high-symmetry adsorption sites (\textit{top}, hollow \textit{fcc}, hollow \textit{hcp}) in the space between the graphene layer and the Ni(111) surface. In this structure there are four different occupation sites for carbon atoms in the unit cell with respect to the adsorption sites of the Ni(111) surface when they are placed above a Ni or Al atom. They are labeled in Fig.~\ref{fig:structure} as $\mathrm{C}_\mathrm{Ni}^{top}$,  $\mathrm{C}_\mathrm{Ni}^{fcc}$,  $\mathrm{C}_\mathrm{Al}^{top}$,  and $\mathrm{C}_\mathrm{Al}^{fcc}$, respectively. A mean distance between the graphene and Al layer is $3.312$\,\AA, indicating ``weak'' bonded graphene-metal system~\cite{Voloshina:2011NJP}. The latter fact can be evaluated when considering the electronic band structure of the system under study: all electronic bands of graphene are shifted to lower binding energies compared to graphene/Ni(111). Furthermore, the electronic structure of the graphene layer as well as the Dirac cone in the vicinity of $E_F$ are fully restored (there is a small electron doping of graphene leading to a shift of the Dirac point below $E_F$ by ca.\,$0.64$\,eV). Thus, intercalation of Al decouples the electronic structure of graphene from the substrate, preventing thereby hybridization between graphene $\pi$ and Ni $3d$ valence band states. In the present system there is no strong hybridization between graphene and Al valence band states conserving the Dirac cone. There is only one energy region where such hybridization is visible: around the $M$ point at approximately $E -E_F= 1$\,eV (see Ref.~\cite{Voloshina:2011NJP}).

The decoupling process in this case is also easily visible when considering the NEXAFS spectra obtained for the system of interest. Intercalation of a thin Al layer underneath a graphene layer on Ni(111) leads to drastic changes in the absorption spectrum compared to the one for graphene/Ni(111): The shape of the spectrum, positions of main spectroscopic features as well as the energy separation between $\pi^*$ and $\sigma^*$ features become similar to those in the spectrum of free-standing graphene [Fig.~\ref{fig:theoryvsexp} (lower panel)]. Due to the avoiding hybridisation between the valence band states of graphene and metal in this case, the individual spectra evaluated for the C-atoms of the four types are indistinguishable from each other  (see Supplementary Fig. S3~\cite{suppl}). Weak hybridisation between graphene $\pi^*$ and Al $3p$ states does not affect the result. Note: agreement between theory and experiment is again excellent.

\subsection{Bilayer-graphene/Ni(111)}

The energetically most stable arrangement of the graphene layers above Ni(111) in this case corresponds to the situation when carbon atoms of the first layer are adopting the \textit{top-fcc} configuration [as it is in the case of the single graphene layer deposited on Ni(111)] and the second graphene layer is displaced horizontally with respect to the first graphene layer as it is in bulk graphite in the way, that carbon atoms from the graphene unit cell are occupying the both hollow sites with respect to the Ni(111) surface [Fig.~\ref{fig:structure} (right-hand side figure)]. The resulting configuration can therefore be named as \textit{top-fcc/fcc-hcp} (see Supplementary Tab. S1~\cite{suppl}).

\begin{figure}
\centering
  \includegraphics[width=0.96\textwidth]{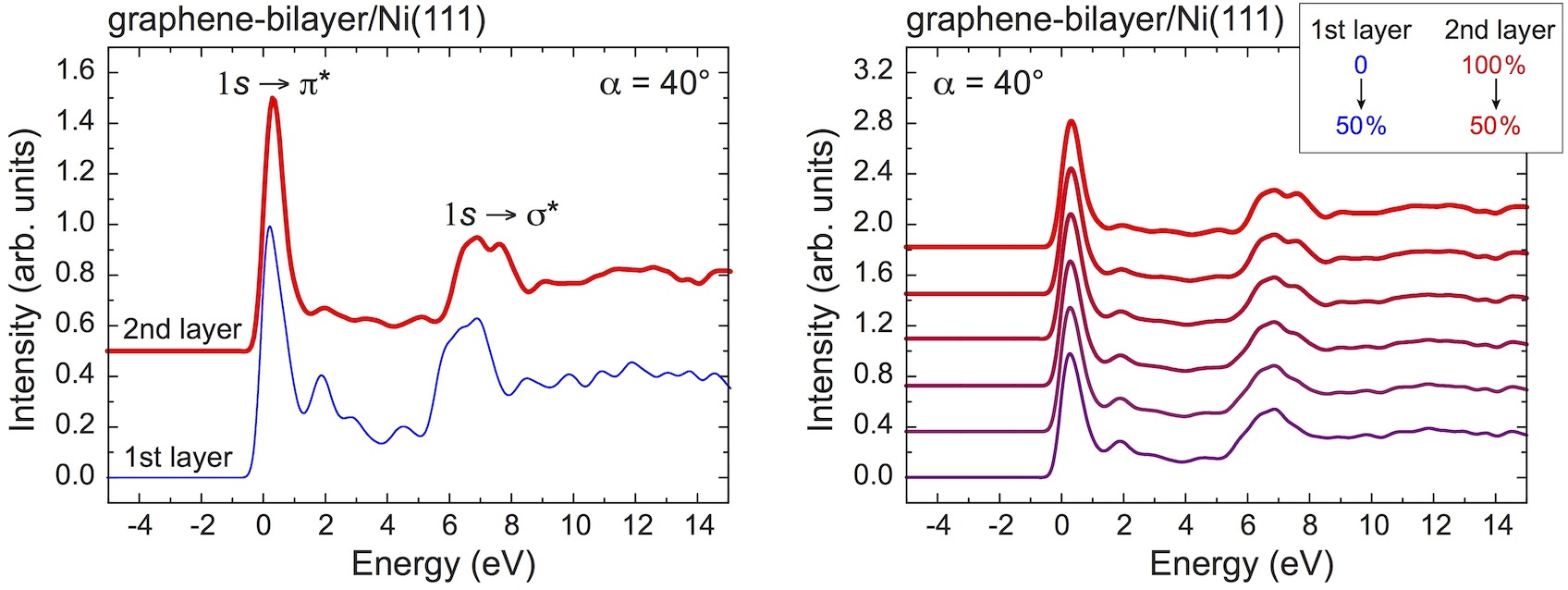}
  \caption{(Left panel) Calculated C $K$ NEXAFS spectra for $\alpha = 40^\circ$ for the graphene-bilayer/Ni(111) system. Spectra are separated on the contributions from the 1st and 2nd layer, respectively. (Right panel) Sequence of the simulated C $K$ NEXAFS spectra for graphene-bilayer/Ni(111), where the contribution from both layers is increased/decreased from top to bottom in order to model the change of surface-vs-bulk sensitivity in experimental data.}
  \label{fig:bilayer}
\end{figure}

In the studied system, the interaction energy between the top (second) graphene layer and graphene/Ni(111) is $91$\,meV per graphene unit cell that is comparable with the value known for the interlayer interaction in graphite ($122\pm 10$\,meV~\cite{Zacharia:2004}). Therefore we can expect, that bilayer-graphene/Ni(111) should display properties of the both systems, free-standing graphene and graphene/Ni(111). When looking at the results of the electronic band structure calculations, one can see that, indeed, while valence band states of the first graphene layer and Ni are strongly hybridized and Dirac cone is destroyed (see Supplementary Fig. S4), picture observed for the second graphene layer is much more similar to that of the free-standing graphene: there is no hybridisation visible and Dirac cone is preserved (see Supplementary Fig. S5~\cite{suppl}). At the same time, a band gap of ca.\,$0.15$\,eV is opened at the $K$-point at $E-E_F=-0.5$\,eV. There is, however, no significant difference between pictures observed for the empty states of the second graphene layer and those of the free-standing graphene. The same is true for the empty states of the first graphene layer in this system and graphene in graphene/Ni(111). This is reflected in the NEXAFS spectra calculated for the bilayer: while results obtained for the first graphene layer are identical with those of graphene/Ni(111), the NEXAFS spectra of the second layer are similar to those obtained for the free-standing graphene as well as graphene in the graphene/Al/Ni(111) trilayer [Fig.~\ref{fig:bilayer} (left panel)].

As discussed earlier the NEXAFS spectra can be collected in two modes, more bulk-sensitive TEY and more surface-sensitive PEY modes. Here, the variation of the repulsive potential (which suppress the secondary electron tail) applied to the electrode before electron detector allows for the gradual variation of the sensitivity (surface vs bulk). Fig.~\ref{fig:bilayer} (right panel) shows the NEXAFS spectra of graphene-bilayer/Ni(111) where the sensitivity is varied from TEY mode (lower spectrum) to PEY mode (upper spectra). The clear change of the shape of the spectra to the one characteristic for the free-standing graphene is visible. These simulations allow to discriminate between single and bilayer graphene in the investigated system.  

\section{Conclusions}

Here we presented the theoretical description of the \mbox{NEXAFS} spectra of the graphene layers on metallic substrate. The interaction in these systems is varied from ``strong'' (when the electronic structure of graphene in the vicinity of $E_F$ is strongly modified) to ``weak'' (when the electronic structure of the graphene layer is a free-standing-graphene like). All these variations in interaction and in electronic structure are consequently reflected in the simulated NEXAFS spectra which can be used as a fingerprint for the description of interaction at the graphene/metal interface. The presented results are compared with the available experimental data and they show the excellent agreement between theory and experiment.

\section*{Acknowledgements}
We appreciate the support from the German Academic Exchange Service (DAAD) through the Center of Excellence: ``German-Russian Interdisciplinary Science Center (G-RISC)''. The High Performance Computing Network of Northern Germany (HLRN) is acknowledged for computer time.

\clearpage

\clearpage

\clearpage

\clearpage

\clearpage

\clearpage

\clearpage

\noindent
Supplementary material for the manuscript:\\

\noindent\textbf{Theoretical description of X-ray absorption spectroscopy of the graphene-based systems}\\
\newline
Elena Voloshina,\textit{$^{a}$} Roman Ovcharenko,\textit{$^{b}$} Alexander Shulakov,\textit{$^{b}$} and Yuriy Dedkov\textit{$^{c}$}\\
\newline
\small
\textit{$^{a}$Humboldt-Universit\"at zu Berlin, Institut f\"ur Chemie, 12489 Berlin, Germany}\\
\textit{$^{b}$\mbox{V.~A.~Fock Institute of Physics, Saint-Petersburg State University, 198405 St. Petersburg, Russia}}\\
\textit{$^{c}$SPECS Surface Nano Analysis GmbH, Voltastra\ss e 5, 13355 Berlin, Germany}
\newline
\newline
\normalsize
\noindent \textbf{Content:}
\begin{itemize}

\item[1.]  Description of the calculation algorithm of the NEXAFS spectra.

\item[2.]  Table S1:  Results for the graphene/Ni(111) and graphene-bilayer/Ni(111). All adsorption energies (AE) are given per graphene unit cell. 

\item[3.]  Figure S1: Calculated majority and minority spin band structures (black) for a slab terminated by graphene/Ni(111) interface (\textit{top-fcc} arrangement). The bands replotted in red (blue) using the carbon $p_z$ ($sp^2$) character as a weighting factor are superimposed. Contributions from the $\mathrm{C}^{top}$ and $\mathrm{C}^{fcc}$ are separated.

\item[4.]  Figure S2: Calculated C $K$ NEXAFS spectra for $\alpha = 40^\circ$ for the graphene/Ni(111) system. Spectra are separated on the contributions from the $\mathrm{C}^{top}$ and $\mathrm{C}^{fcc}$, respectively.

\item[5.]  Figure S3: Calculated C $K$ NEXAFS spectra for $\alpha = 40^\circ$ for the graphene/Al/Ni(111) system. Spectra are separated on the contributions from the $\mathrm{C}^{top}_\mathrm{Al}$, $\mathrm{C}^{top}_\mathrm{Ni}$, $\mathrm{C}^{fcc}_\mathrm{Al}$, and $\mathrm{C}^{fcc}_\mathrm{Ni}$, respectively.

\item[6.]  Figure S4: Calculated majority and minority spin band structures (black) for a slab terminated by graphene-bilayer/Ni(111) interface (\textit{top-fcc/fcc-hcp} arrangement). The bands replotted in red (blue) using the carbon $p_z$ ($sp^2$) character of the first graphene layer as a weighting factor are superimposed. Contributions from the $\mathrm{C}^{top}$ and $\mathrm{C}^{fcc}$ are separated.

\item[7.]  Figure S5: Calculated majority and minority spin band structures (black) for a slab terminated by graphene-bilayer/Ni(111) interface (\textit{top-fcc/fcc-hcp} arrangement). The bands replotted in red (blue) using the carbon $p_z$ ($sp^2$) character of the second graphene layer as a weighting factor are superimposed. Contributions from the $\mathrm{C}^{hcp}$ and $\mathrm{C}^{fcc}$ are separated.

\item[8.]  Figure S6: Calculated C $K$ NEXAFS spectra for $\alpha = 40^\circ$ for the free-standing graphene as obtained by means of different approximations (SK = Slater-Koster method).

\item[9.]  Figure S7: Calculated C $K$ NEXAFS spectra for $\alpha = 40^\circ$ for the graphene/Ni(111) system as obtained by means of different approximations (SK = Slater-Koster method).

\item[10.]  Figure S8: Calculated C $K$ NEXAFS spectra for $\alpha = 40^\circ$ for the graphene/Al/Ni(111) system as obtained by means of different approximations (SK = Slater-Koster method).

\item[11.]  Figure S9: Calculated C $K$ NEXAFS spectra for $\alpha = 40^\circ$ for the graphene-bilayer/Ni(111) system as obtained by means of different approximations (SK = Slater-Koster method).

\end{itemize}

\linespread{1.2}

\clearpage

\noindent
\textbf{Description of the calculation algorithm of the NEXAFS spectra.}
\newline
\newline
Let us consider the radiative X-ray transitions of an electron from the core level $c \equiv \{n_c, l_c, m_c\}$ on the band states $n \mathbf{k}$ localized on the atom $A$. In the tight-binding approximation, neglecting the ``cross'' transition contribution to the intensity of radiation, the NEXAFS intensity $I(E)$ can be represented as the sum of the products of the electric dipole probabilities of transition $W_{\mu}$ from the core level on the atom-like valence band state $\mu$ with partial density of states $N_{\mu}(E)$
\begin{equation}
I(E) \sim \sum_{\mu} W_{\mu} \, N_{\mu}(E) \,,
\end{equation}
\begin{equation}
W_{\mu} \sim \sum_{m_c, \alpha} |A^{\alpha}_{\mu m_c}|^2 \,,
\label{probab1}
\end{equation}
where the amplitude $A^{\alpha}_{\mu m_c}$ of transition is given by
\begin{equation}
A^{\alpha}_{\mu m_c}=\langle \varphi_{\mu}|r_{\alpha} |\varphi_c \rangle \,.
\end{equation}
Here $\varphi_{\mu}$ and $\varphi_c$ are the atomic wave functions of  valence electron and core hole, respectively. In the framework of the one-particle approximation, the partial density of states (PDOS) can be expressed via the imaginary part of the advanced Green's function projected onto the atomic orbitals $G^{-}(E)$
\begin{equation}
N_{\mu}(E) = \frac{1}{\pi} \, \mathrm{Im} G^{-}_{\mu \mu}(E)\,.
\label{dens1}
\end{equation}
%
%If to 
Completely ignoring the influence of the core hole, the Green's function can be written as
\begin{equation}
\label{G_final}
G^{-}_{\mu \nu}(E) = \sum_{n, \mathbf{k}} 
\frac{\langle \mu | \psi_{n \mathbf{k}} \rangle \,
\langle \psi_{n \mathbf{k}} | \nu \rangle}
{E-E_n(\mathbf{k})-E_{\rm f}+i \delta}\,,
\end{equation}
with $E_{\rm f}$ -- the Fermi energy, $E_n(\mathbf{k})$ and $\psi_{n \mathbf{k}}$ -- the unperturbed band energy and wave function, respectively. 
It may be considered as an assumption that the lifetime of the core hole is much smaller than the relaxation time of the system (an approximation of the final state).

Another extreme case takes place when the band wave functions are completely changed because of the core hole Coulomb potential (initial state approximation). Then the expression for the Green's function takes the form
\begin{equation}
\label{G_initial}
\tilde G^{-}_{\mu \nu }(E) = \sum_{n, \mathbf{k}} 
\frac{\langle \mu | \tilde \psi_{n \mathbf{k}} \rangle \,
\langle \tilde \psi_{n \mathbf{k}} | \nu \rangle}
{E-\tilde E_n(\mathbf{k})-E_{\rm f}+i \delta} \,,
\end{equation}
where $\tilde E_n(\mathbf{k})$ and $\tilde \psi_{\mathbf{k}}$ are perturbed band energies and wave functions, respectively. For PDOS in the presence of the core hole one gets
\begin{equation}
\tilde N_{\mu}(E) = \frac{1}{\pi} \, \mathrm{Im} \tilde G^{-}_{\mu \mu}(E)\, .
\label{dens2}
\end{equation}

MND theory considers the core hole filling completely in the dynamical way. The expression for the NEXAFS intensity within the MND theory in the form 
\begin{eqnarray}
\label{I_MND}
I_{\rm MND}(E) \sim \mathrm{Re} \sum_{\mu, \nu} W_{\mu} 
\int \limits_{E_{\rm f}}^{\infty} d E^{\prime} \, 
\tilde N_{\mu \nu}(E^{\prime})  
\int\limits_0^{+\infty} dt \, \times %\nonumber \\
e^{i \, (E+E_c-E^{\prime})\,t+\Delta(-t)} \, \tilde\varphi_{\nu \mu }(E^{\prime},t)\,,
\end{eqnarray}
where the matrix $\tilde\varphi(E,t)$ satisfies the system of integral equations
\begin{eqnarray}
\tilde\varphi_{\mu \nu}(E,t) = \delta_{\mu \nu} + 
\sum_{\mu^{\prime},\nu^{\prime}} 
V_{\mu \mu^{\prime}} \int\limits_{E_{\rm f}}^{\infty}
dE^{\prime} \, \times %\nonumber \\ 
L_{\mu^{\prime} \nu^{\prime}}(E,E^{\prime},t) \,
\tilde\varphi_{\nu^{\prime} \nu}(E^{\prime},t)\,. 
\end{eqnarray}
The kernel of the integral operator $L_{\mu \nu}(E,E^{\prime},t)$ is given by
\begin{eqnarray}
L_{\mu \nu}(E,E^{\prime},t) = \sum_{\mu^{\prime}}
\frac{I_{\mu \mu^{\prime}}(E^{\prime},t)-e^{i (E-E^{\prime}) t} 
I_{\mu \mu^{\prime}}(E,t)}{E^{\prime}-E} \times %\nonumber \\
\tilde N_{\mu^{\prime} \nu}(E^{\prime})\,,
\end{eqnarray}
where

\begin{eqnarray}
\label{I_fun}
I_{\mu \nu}(E,t) = \delta_{\mu \nu} + \sum_{\mu^{\prime}} V_{\mu \mu^{\prime}}
\int\limits_{-\infty}^{+\infty} d E^{\prime} \,
\frac{e^{i(E-E^{\prime})t}-1}{E-E^{\prime}} \, \times %\nonumber \\
N_{\mu^{\prime} \nu}(E^{\prime}) \,,
\end{eqnarray}
\begin{equation}
\label{Core_potential}
V_{\mu \nu} = \sum_{n, \mathbf{k}} \left( \langle \mu | \psi_{n \mathbf{k}} 
\rangle 
\langle \psi_{n \mathbf{k}} | \nu \rangle E_n(\mathbf{k}) -  
\langle \mu | \tilde \psi_{n \mathbf{k}} \rangle \langle \tilde \psi_{n 
\mathbf{k}} 
| \nu \rangle \tilde E_n(\mathbf{k}) \right)\,.
\end{equation}
In the above expressions, the matrixes $N_{\mu \nu}$ and $\tilde N_{\mu \nu}$ have the form
\begin{equation}
N_{\mu \nu} = \frac{1}{\pi} \, \mathrm{Im} \, G^{-}_{\mu \nu }(E) \,, \qquad
\tilde N_{\mu \nu} = \frac{1}{\pi} \, \mathrm{Im} \, \tilde G^{-}_{\mu \nu }(E) \,.
\end{equation}
Note that the diagonal elements of matrixes $N_{\mu \nu}$ and $\tilde N_{\mu \nu}$ coincide with the PDOS defined above in Eqs.\,(\ref{dens1}) and (\ref{dens2}) in the final and initial state approximations.

There are two ways to define input data for the MND calculations expressed by Eqs.\,(\ref{I_MND})--(\ref{I_fun}). The first approach is to use the expressions (\ref{G_final}), (\ref{G_initial}), and (\ref{Core_potential}) to introduce the Green functions $G^{-}(E), \tilde G^{-}(E)$ and the core potential $V$. It works well if the band wave functions basis set is large enough. If it is not the case, the core hole potential $V$ depends on energy. To minimize the error caused by the energy dependence of the potential one can use a second approach. The unperturbed Green function $G^-(E)$ and core hole potential $V$ defined from equations (\ref{G_initial}) and (\ref{Core_potential}), respectively, whereas the perturbed Green function $\tilde G^-(E)$ is obtained from the Slater-Koster equation
\begin{equation}
\tilde G^-_{\mu \nu}(E) = \sum_{\nu'} G^-_{\mu \nu'}(E) \left[ T^{-1} \right]_{\nu' \nu},
\end{equation}
where
\begin{equation}
T_{\mu \nu} = \delta_{\mu \nu} + \sum_{\nu'} V_{\mu \nu'} G^-_{\nu' \nu}(E)
\end{equation}
Although the potential still depends on energy, the perturbed Green function now takes it into account.

\clearpage

\begin{table*}
\raggedright
\noindent\textbf{Table\,S1.} Results for the graphene/Ni(111) and graphene-bilayer/Ni(111). All adsorption energies (AE) are given per graphene unit cell. \\
\vspace{0.3cm}
\begin{ruledtabular}
\begin{tabular}{llll}
System & $AE_1$ (eV)&$AE_2$ (eV)&$AE_3$ (eV)\\
\hline
%\textbf{graphite}&$-0.13368$&&\\
%&&&\\			
%\multicolumn{4}{l}{\textbf{graphene-bilayer}}\\			
%\quad AB-stacking&$-0.04969$&&\\		
%\quad AA-stacking&$-0.03707$&&\\
%&&&\\		
\multicolumn{4}{l}{\textbf{graphene/Ni(111)}}\\			
\quad \textit{top-fcc}				&&$-0.284$& \\		
\quad \textit{top-hcp}				&&$-0.257$& \\		
\quad \textit{fcc-hcp}				&&$-0.060$& \\	
%&&&\\		
%\textbf{graphene/Al/Ni(111)}&$-0.08718$&&\\			
&&&\\		
\multicolumn{4}{l}{\textbf{graphene-bilayer/Ni(111)}}\\			
\quad \textit{top-fcc~/~top-hcp}		&$-0.091$&$-0.187$&$-0.275$\\
\quad \textit{top-fcc~/~fcc-hcp}		&$-0.096$&$-0.190$&$-0.280$\\
\quad \textit{top-fcc~/~top-fcc}		&$-0.067$&$-0.175$&$-0.277$\\
&&&\\		
%Gr$_2$/Ni(111)&&&\\	
\quad \textit{top-hcp~/~top-fcc}	&$-0.089$&$-0.173$&$-0.247$\\
\quad \textit{top-hcp~/~fcc-hcp}	&$-0.096$&$-0.176$&$-0.253$\\
\quad \textit{top-hcp~/~top-hcp}	&$-0.067$&$-0.162$&$-0.250$\\
&&&\\	
%Gr$_2$/Ni(111)&&&\\			
\quad \textit{fcc-hcp~/~top-hcp} 	&$-0.128$&$-0.094$&$-0.088$\\
\quad \textit{fcc-hcp~/~top-fcc}   	&$-0.140$&$-0.100$&$-0.101$\\
\quad \textit{fcc-hcp~/~fcc-hcp} 	&$-0.101$&$-0.080$&$-0.087$\\
\end{tabular}
\end{ruledtabular}\\
\vspace{0.5cm}
$AE_1=E_\mathrm{gr_2/Ni(111)}-E_\mathrm{gr/Ni(111)}-E_\mathrm{gr}$\\
$AE_2=E_\mathrm{gr_n/Ni(111)}-E_\mathrm{Ni(111)}-E_\mathrm{gr}\times n$\quad ($n$ - number of graphene layers)\\
$AE_3=E_\mathrm{gr_2/Ni(111)}-E_\mathrm{Ni(111)}-E_\mathrm{gr_2}$

\end{table*}

\clearpage

\begin{figure}
\centering
\includegraphics[width=0.8\textwidth]{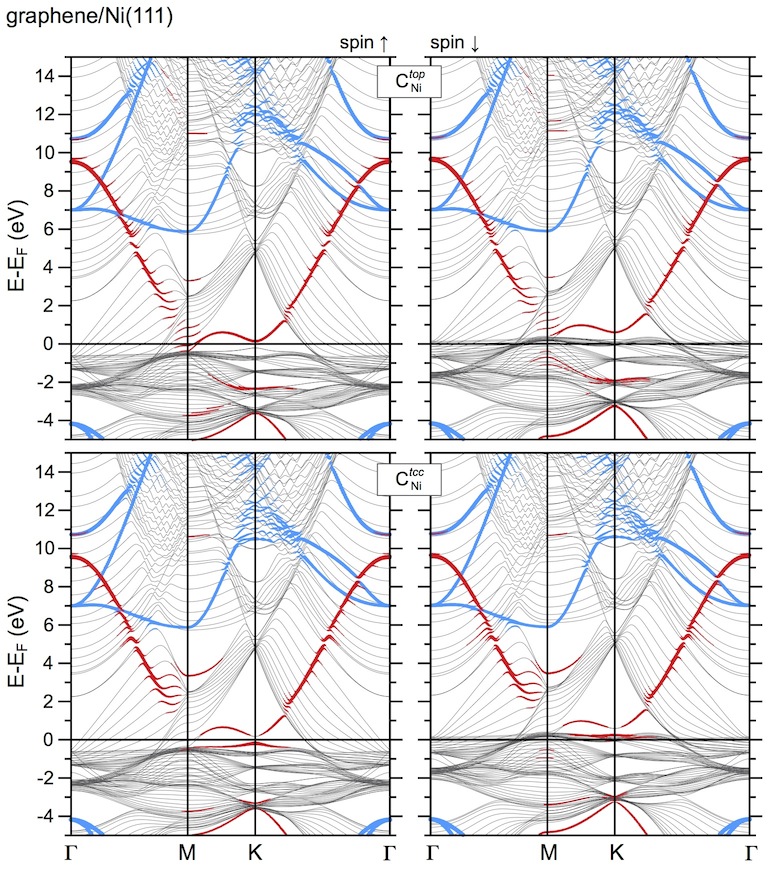}
\end{figure}
\noindent\textbf{Figure\,S1.} Calculated majority and minority spin band structures (black) for a slab terminated by graphene/Ni(111) interface (\textit{top-fcc} arrangement). The bands replotted in red (blue) using the carbon $p_z$ ($sp^2$) character as a weighting factor are superimposed. Contributions from the $\mathrm{C}^{top}$ and $\mathrm{C}^{fcc}$ are separated.

\clearpage

\begin{figure}
\centering
\includegraphics[width=0.8\textwidth]{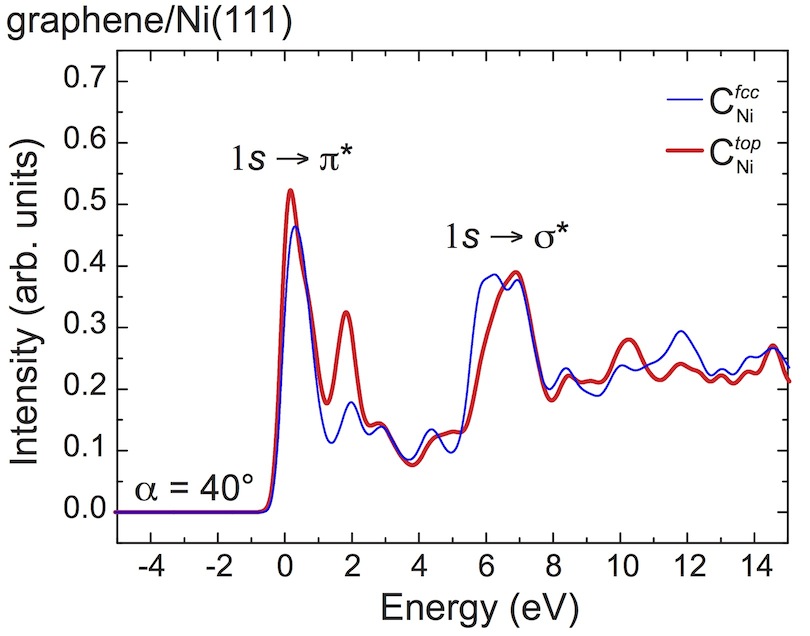}
\end{figure}
\noindent\textbf{Figure\,S2.}  Calculated C $K$ NEXAFS spectra for $\alpha = 40^\circ$ for the graphene/Ni(111) system. Spectra are separated on the contributions from the $\mathrm{C}^{top}$ and $\mathrm{C}^{fcc}$, respectively.

\clearpage

\begin{figure}
\centering
\includegraphics[width=0.8\textwidth]{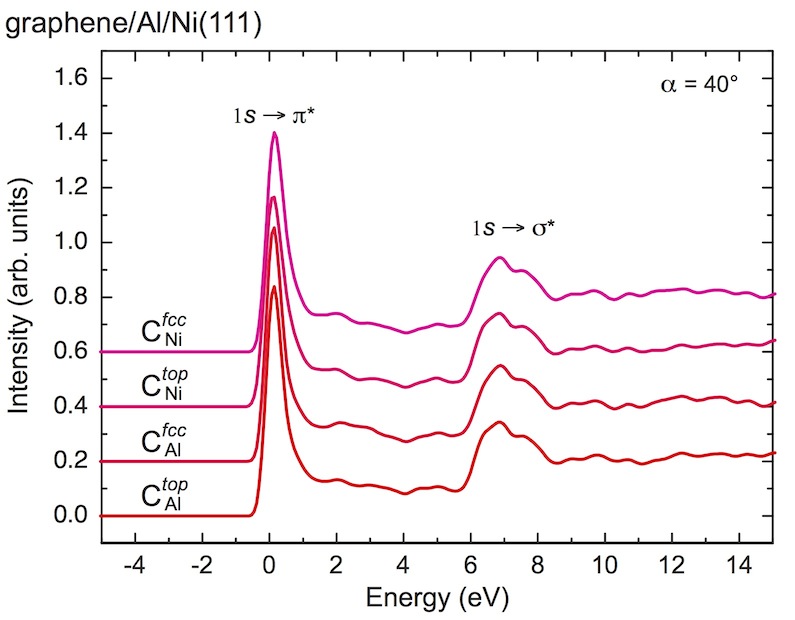}
\end{figure}
\noindent\textbf{Figure\,S3.}  Calculated C $K$ NEXAFS spectra for $\alpha = 40^\circ$ for the graphene/Al/Ni(111) system. Spectra are separated on the contributions from the $\mathrm{C}^{top}_\mathrm{Al}$, $\mathrm{C}^{top}_\mathrm{Ni}$, $\mathrm{C}^{fcc}_\mathrm{Al}$, and $\mathrm{C}^{fcc}_\mathrm{Ni}$, respectively.
\clearpage

\begin{figure}
\centering
\includegraphics[width=0.8\textwidth]{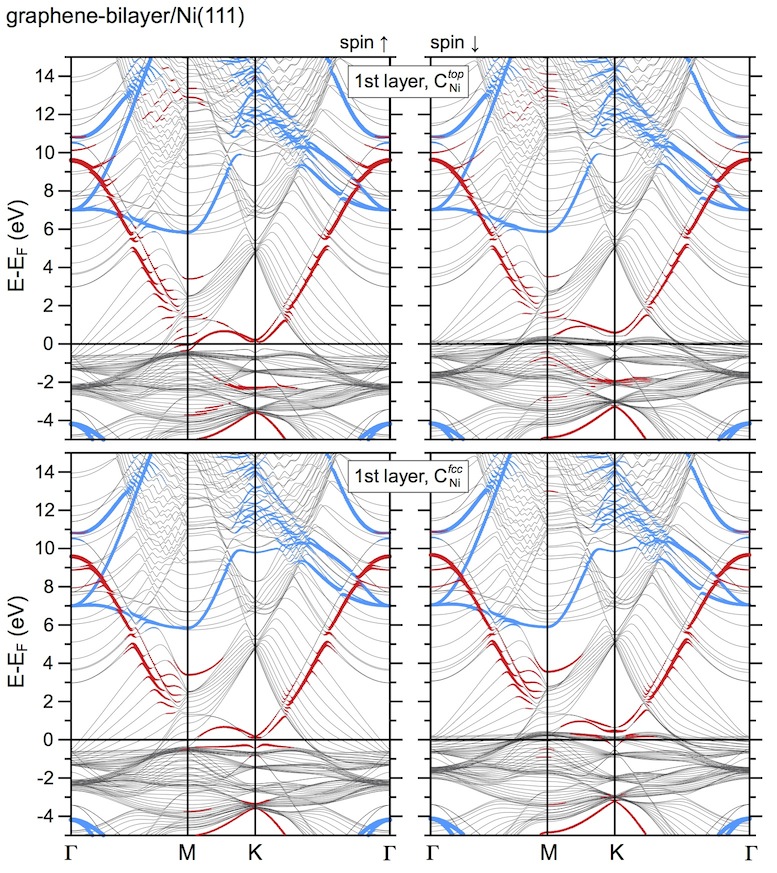}
\end{figure}
\noindent\textbf{Figure\,S4.} Calculated majority and minority spin band structures (black) for a slab terminated by graphene-bilayer/Ni(111) interface (\textit{top-fcc/fcc-hcp} arrangement). The bands replotted in red (blue) using the carbon $p_z$ ($sp^2$) character of the first graphene layer as a weighting factor are superimposed. Contributions from the $\mathrm{C}^{top}$ and $\mathrm{C}^{fcc}$ are separated.

\clearpage

\begin{figure}
\centering
\includegraphics[width=0.8\textwidth]{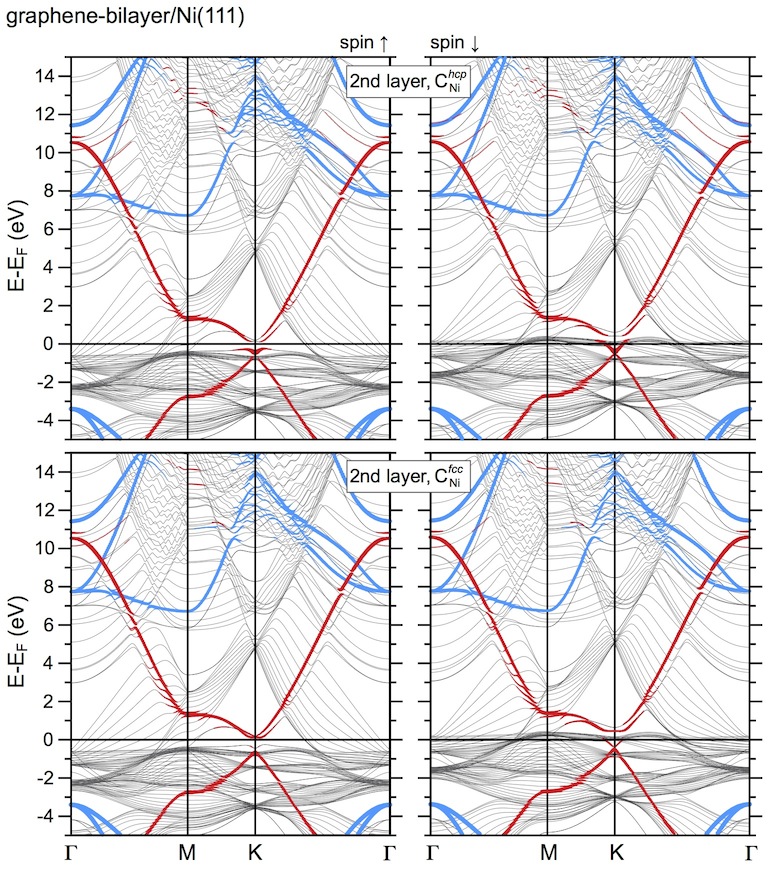}
\end{figure}
\noindent\textbf{Figure\,S5.} Calculated majority and minority spin band structures (black) for a slab terminated by graphene-bilayer/Ni(111) interface (\textit{top-fcc/fcc-hcp} arrangement). The bands replotted in red (blue) using the carbon $p_z$ ($sp^2$) character of the second graphene layer as a weighting factor are superimposed. Contributions from the $\mathrm{C}^{hcp}$ and $\mathrm{C}^{fcc}$ are separated.

\clearpage

\begin{figure}
\centering
\includegraphics[width=0.8\textwidth]{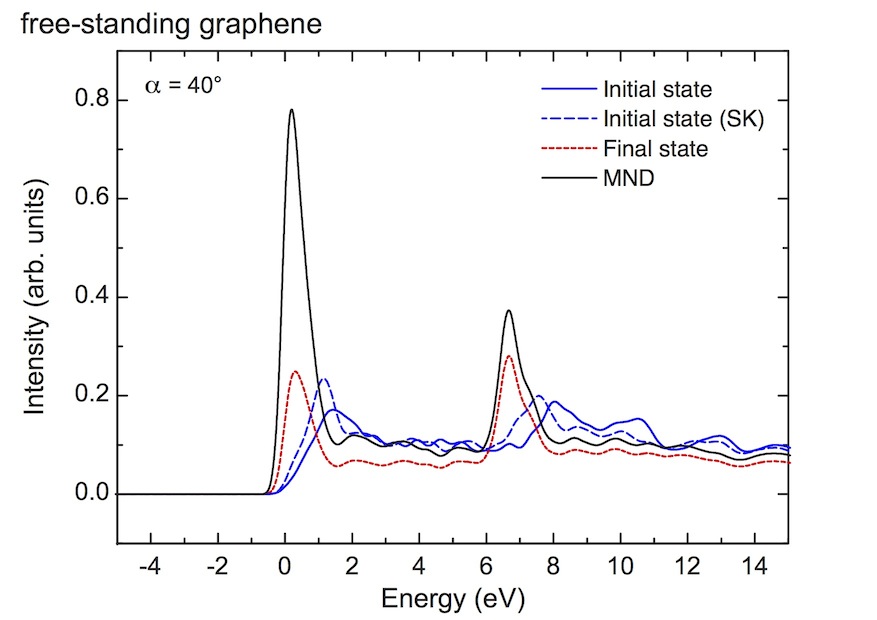}
\end{figure}
\noindent\textbf{Figure\,S6.} Calculated C $K$ NEXAFS spectra for $\alpha = 40^\circ$ for the free-standing graphene as obtained by means of different approximations (SK = Slater-Koster method).

\clearpage

\begin{figure}
\centering
\includegraphics[width=0.8\textwidth]{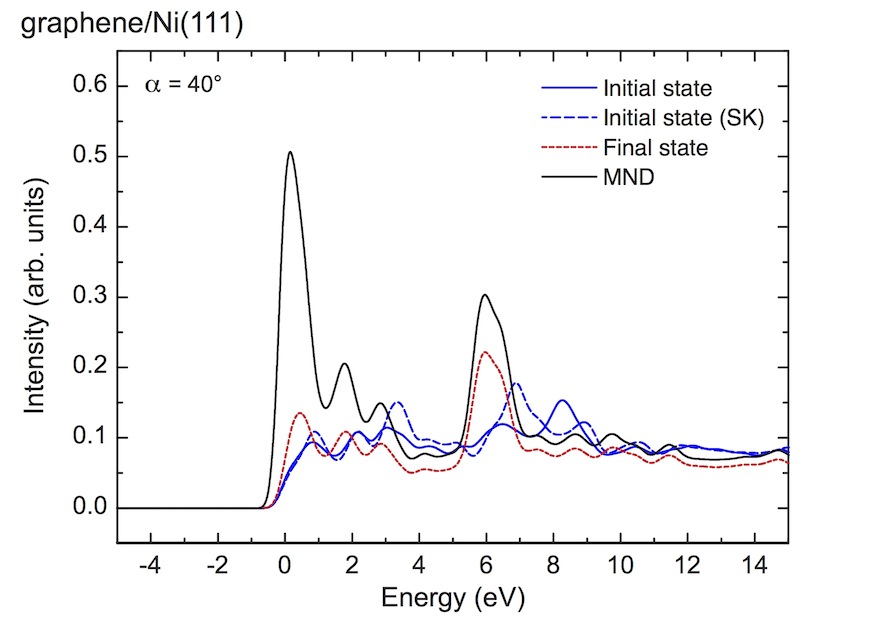}
\end{figure}
\noindent\textbf{Figure\,S7.} Calculated C $K$ NEXAFS spectra for $\alpha = 40^\circ$ for the graphene/Ni(111) system as obtained by means of different approximations (SK = Slater-Koster method).

\clearpage

\begin{figure}
\centering
\includegraphics[width=0.8\textwidth]{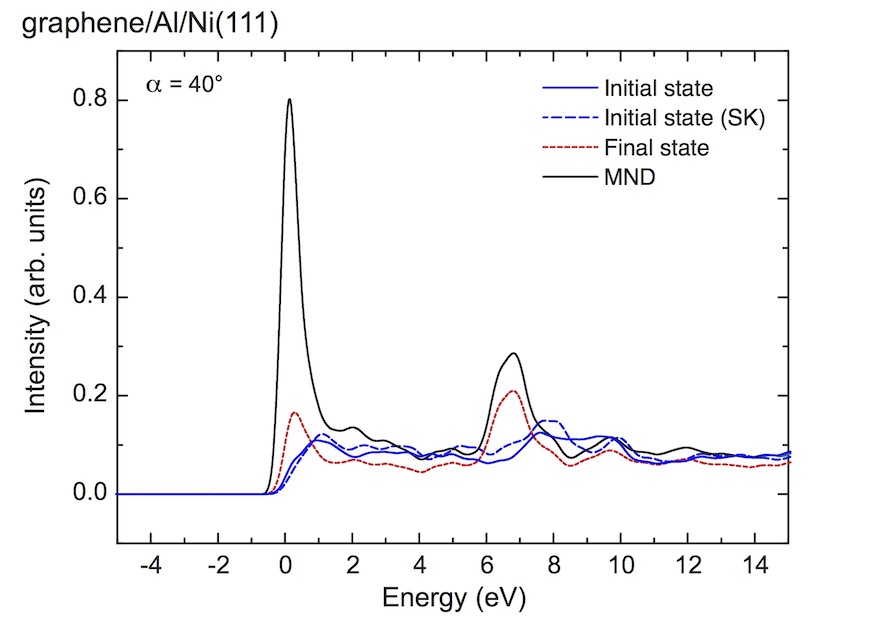}
\end{figure}
\noindent\textbf{Figure\,S8.} Calculated C $K$ NEXAFS spectra for $\alpha = 40^\circ$ for the graphene/Al/Ni(111) system as obtained by means of different approximations (SK = Slater-Koster method).

\clearpage

\begin{figure}
\centering
\includegraphics[width=0.8\textwidth]{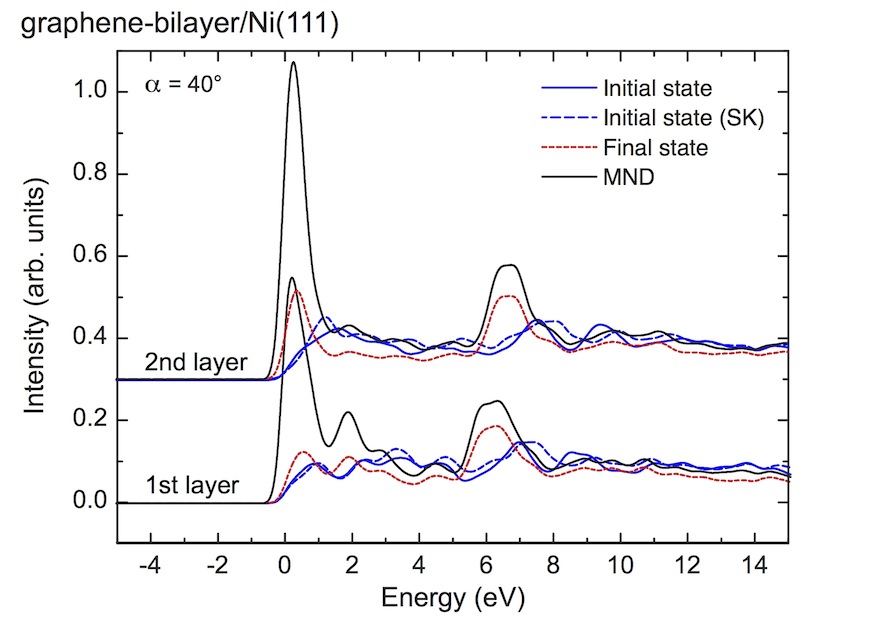}
\end{figure}
\noindent\textbf{Figure\,S9.} Calculated C $K$ NEXAFS spectra for $\alpha = 40^\circ$ for the graphene-bilayer/Ni(111) system as obtained by means of different approximations (SK = Slater-Koster method).


\begin{thebibliography}{99}

\bibitem{Stohr:1992NEXAFSbook}
J.~St\"ohr, \emph{NEXAFS Spectroscopy}, Springer, Berlin, 1992.
 
\bibitem{Frazer:2003a}
B.~H. Frazer, B.~Gilbert, B.~R. Sonderegger and G.~De~Stasio, Surf. Sci. \textbf{537}, 161--167 (2003).
 
\bibitem{Stohr:1999b}
J.~St\"ohr and M.~Samant, J. Electr. Spectr. Rel. Phenom. \textbf{98}, 189--207 (1999).
 
\bibitem{Stohr:1999a}
J.~St\"ohr, J. Magn. Magn. Mat. \textbf{200}, 470--497 (1999). 
 
\bibitem{Thole:1992}
B.~Thole, P.~Carra, F.~Sette and G.~van~der Laan, Phys. Rev. Lett. \textbf{68}, 1943--1946 (1992).
 
\bibitem{Carra:1993}
P.~Carra, B.~Thole, M.~Altarelli and X.~Wang, Phys. Rev. Lett. \textbf{70}, 694--697 (1993).
 
\bibitem{Bruhwiler:1995}
P.~Br{\"u}hwiler, A.~Maxwell, C.~Puglia, A.~Nilson, S.~Anderson and N.~M\aa rtenson, Phys. Rev. Lett. \textbf{74}, 614--617 (1995).
 
\bibitem{Pacile:2008}
D.~Pacil{\'e}, M.~Papagno, A.~F. Rodr{\'\i}guez, M.~Grioni and L.~Papagno, Phys. Rev. Lett. \textbf{101}, 066806 (2008).
 
\bibitem{Hua:2010}
W.~Hua, B.~Gao, S.~Li, H.~{\AA}gren and Y.~Luo, Phys. Rev. B \textbf{82}, 155433 (2010).
 
\bibitem{Wessely:2005}
O.~Wessely, M.~Katsnelson and O.~Eriksson, Phys. Rev. Lett. \textbf{94}, 167401 (2005).
 
\bibitem{Wessely:2006}
O.~Wessely, O.~Eriksson and M.~Katsnelson, Phys. Rev. B \textbf{73}, 075402 (2006).
 
\bibitem{Wintterlin:2009}
J.~Wintterlin and M.~L. Bocquet, Surf. Sci. \textbf{603}, 1841--1852 (2009).
 
\bibitem{Batzill:2012}
M.~Batzill, Surf. Sci. Rep. \textbf{67}, 83--115 (2012).
 
\bibitem{Dedkov:2012book}
Y.~S. Dedkov, K.~Horn, A.~Preobrajenskij and M.~Fonin, \emph{Graphene Nanoelectronics}, Springer, Berlin, 2012. 
 
\bibitem{Voloshina:2012c}
E.~Voloshina and Y.~Dedkov, Phys. Chem. Chem. Phys. \textbf{14}, 13502--13514 (2012).
 
\bibitem{Geim:2009}
A.~Geim, Science \textbf{324}, 1530--1534 (2009).
 
\bibitem{Kim:2009a}
K.~S. Kim, Y.~Zhao, H.~Jang, S.~Y. Lee, J.~M. Kim, K.~S. Kim, J.-H. Ahn, P.~Kim, J.-Y. Choi and B.~H. Hong, Nature \textbf{457}, 706--710 (2009). 
 
\bibitem{Bae:2010}
S.~Bae, H.~Kim, Y.~Lee, X.~Xu, J.-S. Park, Y.~Zheng, J.~Balakrishnan, T.~Lei, H.~R. Kim, Y.~I. Song, Y.-J. Kim, K.~S. Kim, B.~Ozyilmaz, J.-H. Ahn, B.~H. Hong and S.~Iijima, Nature Nanotech. \textbf{5}, 574--578 (2010). 
 
\bibitem{Karpan:2007}
V.~M. Karpan, G.~Giovannetti, P.~A. Khomyakov, M.~Talanana, A.~A. Starikov, M.~Zwierzycki, J.~v.~d. Brink, G.~Brocks and P.~J. Kelly, Phys. Rev. Lett. \textbf{99}, 176602 (2007). 
 
\bibitem{Dedkov:2010a}
Y.~S. Dedkov and M.~Fonin, New J. Phys. \textbf{12}, 125004 (2010).
 
\bibitem{Blochl:1994}
P.~Bl\"ochl, Phys. Rev. B \textbf{50}, 17953--17979 (1994). 
 
\bibitem{Perdew:1996}
J.~Perdew, K.~Burke and M.~Ernzerhof, Phys. Rev. Lett. \textbf{77}, 3865--3868 (1996). 
 
\bibitem{Kresse:1994}
G.~Kresse and J.~Hafner, J. Phys.: Condens. Matter \textbf{6}, 8245--8257 (1994).
 
\bibitem{Grimme:2010}
S.~Grimme, J.~Antony, S.~Ehrlich and H.~Krieg, J. Chem. Phys. \textbf{132}, 154104 (2010). 
 
\bibitem{Weser:2010}
M.~Weser, Y.~Rehder, K.~Horn, M.~Sicot, M.~Fonin, A.~B. Preobrajenski, E.~N. Voloshina, E.~Goering and Y.~S. Dedkov, Appl. Phys. Lett. \textbf{96}, 012504 (2010).

\bibitem{suppl}
See Supplemental Material at [URL will be inserted by publisher] for technical material and details of analysis.
 
\bibitem{Bertoni:2004}
G.~Bertoni, L.~Calmels, A.~Altibelli and V.~Serin, Phys. Rev. B \textbf{71}, 075402 (2004).
 
\bibitem{Karpan:2008}
V.~M. Karpan, P.~A. Khomyakov, A.~A. Starikov, G.~Giovannetti, M.~Zwierzycki, M.~Talanana, G.~Brocks, J.~v.~d. Brink and P.~J. Kelly, Phys. Rev. B \textbf{78}, 195419 (2008).
 
\bibitem{Weser:2011}
M.~Weser, E.~N. Voloshina, K.~Horn and Y.~S. Dedkov, Phys. Chem. Chem. Phys. \textbf{13}, 7534--7539 (2011).
 
\bibitem{Rusz:2010}
J.~Rusz, A.~B. Preobrajenski, M.~L. Ng, N.~A. Vinogradov, N.~Martensson, O.~Wessely, B.~Sanyal and O.~Eriksson, Phys. Rev. B \textbf{81}, 073402 (2010).
 
\bibitem{Voloshina:2011NJP}
E.~N. Voloshina, A.~Generalov, M.~Weser, S.~B\"ottcher, K.~Horn and Y.~S. Dedkov, New J. Phys. \textbf{13}, 113028 (2011).
 
\bibitem{Zacharia:2004}
R.~Zacharia, H.~Ulbricht and T.~Hertel, Phys. Rev. B \textbf{69}, 155406 (2004).

\end{thebibliography}
\end{document}